\documentclass[twocolumn,amsmath,amssymb,aps,pra,longbibliography,superscriptaddress,floatfix]{revtex4-1}
\usepackage{physics}
\usepackage{units}
\usepackage{amsmath}
\usepackage{cases}
\usepackage{url}
\usepackage{natbib}
\usepackage{textcase}
\usepackage{amssymb}
\usepackage{graphicx}
\usepackage{bm} 

\usepackage{times}
\usepackage{float}
\usepackage{multirow,microtype,color,relsize}
\usepackage{subfigure}
\usepackage[breaklinks=true]{hyperref}
\usepackage[utf8]{inputenc}
\usepackage[english]{babel}
\hypersetup{colorlinks=true,linkcolor=blue,citecolor=blue}
\hypersetup{linktocpage}
\usepackage{CJKutf8}
\usepackage{breakcites}
\usepackage{float}
\usepackage{siunitx}
\usepackage[dvipsnames]{xcolor}
\definecolor{mypine}{RGB}{1, 121, 111}

\def \be {\begin{equation}}
\def \ee {\end{equation}}

\begin{document}
\begin{CJK*}{UTF8}{gbsn}
\title{Conductivity of two-dimensional narrow gap semiconductors subjected to strong Coulomb disorder}

\author{Yi Huang~(黄奕)}
\affiliation{School of Physics and Astronomy, University of Minnesota, Minneapolis, Minnesota 55455, USA}
\email[Corresponding author: ]{huan1756@umn.edu}

\author{Yanjun He}
\affiliation{Department of Physics, The Ohio State University, Columbus, Ohio 43202, USA}

\author{Brian Skinner}
\affiliation{Department of Physics, The Ohio State University, Columbus, Ohio 43202, USA}

\author{B.\,I. Shklovskii}
\affiliation{School of Physics and Astronomy, University of Minnesota, Minneapolis, Minnesota 55455, USA}

\date{\today}

\begin{abstract}

In the ideal disorder-free situation, a two-dimensional band-gap insulator has an activation energy for conductivity equal to half the band gap $\Delta$. But transport experiments usually exhibit a much smaller activation energy at low temperature, and the relation between this activation energy and $\Delta$ is unclear. Here we consider the temperature-dependent conductivity of a two-dimensional insulator on a substrate containing Coulomb impurities, with random potential amplitude $\Gamma \gg \Delta$.  We show that the conductivity generically exhibits three regimes of conductivity, and only the highest-temperature regime exhibits an activation energy that reflects the band gap. At lower temperatures, the conduction proceeds through activated hopping or Efros-Shklovskii variable-range hopping between electron and hole puddles created by the disorder. We show that the activation energy and characteristic temperature associated with these processes steeply collapse near a critical impurity concentration. Larger concentrations lead to an exponentially small activation energy and exponentially long localization length, which in mesoscopic samples can appear as a disorder-induced insulator-to-metal transition. We also arrive at a similar steep disorder driven insulator-metal transition in thin films of three-dimensional topological insulators with large dielectric constant, for which Coulomb impurities inside the film create a large disorder potential due to confinement of their electric field inside the film.

\end{abstract}
\maketitle
\end{CJK*}

\section{Introduction}
\label{sec:intro}

In a band gap insulator, charged impurities often play a decisive role in determining the properties of the insulating state. Due to the long-range nature of the Coulomb potential that they create, such impurities produce large band bending that changes qualitatively the nature of electron conduction relative to the ideal disorder-free situation. An illustrative case is that of a three-dimensional completely-compensated semiconductor, for which positively charged donors and negatively charged acceptors are equally abundant and randomly distributed in space. In this case, the impurity potential has large random fluctuations, 
which can be screened only when the amplitude of this potential $\Gamma$ reaches $\Delta$, where $2\Delta$ is the band gap. This screening is produced by sparse electron and hole droplets, concentrated in spatially alternating electron and hole puddles~\cite{shklovskii1972,shklovskii1984,skinner2012}. At high enough temperatures, the electrical conductivity is due to activation of electrons and holes from the chemical potential to the energy associated with classical percolation across the sample. 
At lower temperatures, the conductivity is due to hopping between nearest-neighbor puddles. At even smaller temperatures, it is due to variable-range hopping between puddles. Crucially, in each of these temperature regimes, the naive relation $E_a = \Delta$ is lost, where $E_a$ is the activation energy for conductivity. Only in the highest-temperature regime is there a direct proportionality between $E_a$ and $\Delta$ (with a nontrivial small numeric prefactor) \cite{skinner2012, Chen2016}; at lower temperatures, the observed activation energy is nonuniversal and disorder dependent~\cite{shklovskii1972,shklovskii1984}.

In this paper, we consider a similar problem in two dimensions, focusing on the case of strong disorder, $\Gamma \gg \Delta$.
Specifically, we consider a two-dimensional small band-gap semiconductor resting on a thick substrate with a three-dimensional concentration of randomly positioned impurities. We derive the temperature dependence of the electrical conductivity across all temperature regimes and show that observed activation energy can be very small.

Understanding the relation between the energy gap and the observed activation energy for transport is of crucial importance for studying a variety of two-dimensional (2D) electron systems.
For example, recent studies of
2D topological insulators (TIs) \cite{Olshanetsky2015, Kvon2020, Pan2020}, films of 3D TIs~\cite{nandi2018,chong2021,checkelsky2012,chang2013,he2013,mogi2015,liguozhang2017,wang2018,fox2018,moon2019,rosen2019,rodenbach2021,fijalkowski2021}, bilayer graphene (BLG) with an orthogonal electric field~\cite{zou2010,taychatanapat2010}, and twisted bilayer graphene (TBG)~\cite{Serlin_Intrinsic_2020, Stepanov_untying_2020, Park_Flavour_2021, cao_correlated_2018, Cao_unconventional_2018} use the transport activation energy as a way of characterizing small energy gaps. In all these cases, the observed activation energy is much smaller than the energy gap that is expected theoretically or measured through local probes such as optical absorption or scanning tunneling microscopy. 

Here, we show that there is indeed no simple proportionality between the energy gap and the activation energy except at the highest-temperature regime, which is likely irrelevant for many experimental contexts. Instead, we find a wide regime of temperature and disorder strength for which the activation energy is parametrically smaller than the energy gap. At the lowest temperatures the conductivity follows a Efros-Shklovskii (ES) law~\cite{efros1975} rather than an Arrhenius law, and this dependence can give the appearance of a small activation energy.

Let us dwell on two likely applications of our theory. First, our results may be especially relevant for ongoing efforts to understand the energy gaps arising in TBG at certain commensurate fillings of the moir\'{e} superlattice \cite{Serlin_Intrinsic_2020, Stepanov_untying_2020, Park_Flavour_2021, cao_correlated_2018, Cao_unconventional_2018}. Such gaps apparently arise from electron-electron interactions, but the observed activation energies of the maximally insulating state are typically an order of magnitude smaller than the naive interaction scale (see, e.g., Refs.~\onlinecite{Stepanov_untying_2020, Park_Flavour_2021}), and they vary significantly from one sample to another. Scanning tunneling microscopy studies also suggest a gap of the order of ten times larger than the observed activation energy \cite{Xie_spectroscopic_2019, choi2021interactiondriven}. The theory we present here offers a natural way to interpret this discrepancy.

Second, our theory can be applied to the huge body of experimental work on thin films of a 3D TI,
where the surface electrons have a small gap $2\Delta$ due to hybridization of the surface states of two surfaces~\cite{nandi2018,chong2021} or due to intentionally introduced magnetic impurities~\cite{checkelsky2012,chang2013,he2013,mogi2015,liguozhang2017,wang2018,fox2018,moon2019,rosen2019,rodenbach2021,fijalkowski2021}. Understanding the origin of the small apparent activation energy $E_a \ll \Delta$ is crucial for achieving metrological precision of the quantum anomalous Hall effect~\cite{yu2010,chang2013,jinsongzhang2013,mogi2015,fox2018,rodenbach2021,fijalkowski2021} and the quantum spin Hall effect~\cite{liu2010,lu2010,linder2009,chong2021}.

\begin{figure}[t]
    \centering
    \includegraphics[width=\linewidth]{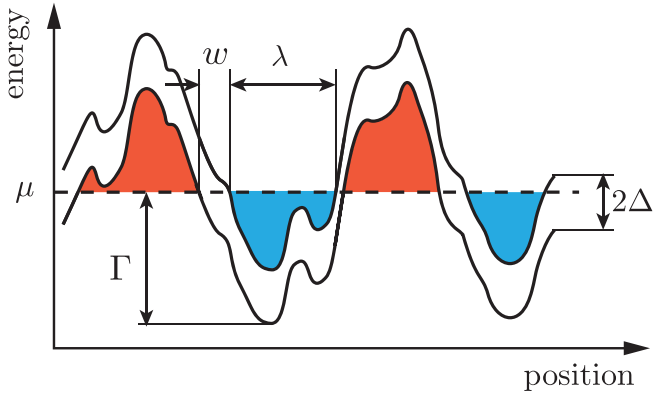}
    \caption{Schematic picture of a cross section of puddles in the case $\Gamma \gg \Delta$. The wavy lines show the conduction-band bottom and the valence-band ceiling separated by the gap $2\Delta$ . The red shaded region above the chemical potential $\mu=0$ represents a hole puddle, while the blue shaded region below $\mu$ represents an electron puddle; $\Gamma$ is the amplitude of the disorder potential, $\lambda$ is the screening length, and $w$ is the width of the barrier between neighboring puddles. } 
    \label{fig:puddles}
\end{figure}

The model we consider of is a two-dimensional semiconductor with band gap $2\Delta$ atop a substrate with a three-dimensional concentration $N$ of random sign charged impurities. 
We assume that the semiconductor has a gapped Dirac dispersion law,
\begin{align} \label{eq:spectrum}
\epsilon^2(\vb{k}) = (\hbar v k)^2 + \Delta^2,
\end{align}
where $\epsilon$ is the electron energy, $\vb{k}$ is the 2D wave vector, $v$ is the Dirac velocity, and $\hbar$ is the reduced Planck constant.
We are interested in the case when the amplitude $\Gamma$ of spatial fluctuations of the random potential satisfies
$\Gamma \gg \Delta$, so that electron and hole puddles occupy almost half of the space each and are separated by a small insulating gap which occupies only a small fraction of space (see Fig.~\ref{fig:puddles}). 
This system is an insulator because in 2D neither electron nor hole puddles percolate, and they are disconnected from each other (neglecting, for now, the possibility of quantum tunneling between puddles). Throughout this paper, we focus on the case of zero chemical potential, for which electron and hole puddles are equally abundant and the system achieves its maximally insulating state.
We argue that this situation is likely realized in the experiments of Refs.~\onlinecite{
Olshanetsky2015, Kvon2020, Pan2020, zou2010, taychatanapat2010, Serlin_Intrinsic_2020, Stepanov_untying_2020, Park_Flavour_2021, cao_correlated_2018, Cao_unconventional_2018, Xie_spectroscopic_2019, choi2021interactiondriven,  nandi2018,chong2021,checkelsky2012,chang2013,he2013,mogi2015,liguozhang2017,wang2018,fox2018,moon2019,rosen2019,rodenbach2021,fijalkowski2021}.

The remainder of this paper is organized as follows. In the following section, we first summarize our main results for the temperature-dependent conductivity. In Sec.~\ref{sec:fractal}, we review the fractal geometry of two-dimensional puddles for the case $\Gamma \gg \Delta$. In Sec.~\ref{sec:tunneling}, we calculate the action accumulated by electrons tunneling across the gap between two neighboring puddles,
the corresponding localization length, and the critical value $(\Gamma/\Delta)_c$ of the ratio $\Gamma/\Delta$, at which crossover to ``almost metallic conductivity'' takes place. 
In Sec.~\ref{sec:hopping}, we calculate the hopping conductivity at $1 \ll \Gamma/\Delta \ll (\Gamma/\Delta)_c$ . 
Section \ref{sec:TI} deals with the generalization of our results to thin TI films. 
Because of the intense recent interest in such films~\cite{liu2010,lu2010,linder2009,zhang2010a,sakamoto2010,zhang2013,kim2013,nandi2018,chong2021,chen2010,yu2010,xu2012,checkelsky2012,jinsongzhang2013,he2013,mogi2015,ye2015,liguozhang2017,wang2018,fox2018,moon2019,rosen2019,tokura2019,deng2020,rodenbach2021,lu2021,fijalkowski2021}, in this section we add a fair amount of numerical estimates.
We close in Sec.~\ref{sec:conclusion} with a summary and conclusion.

\section{Summary of results}
\label{sec:summary}

\begin{figure}[t]
    \centering
    \includegraphics[width=\linewidth]{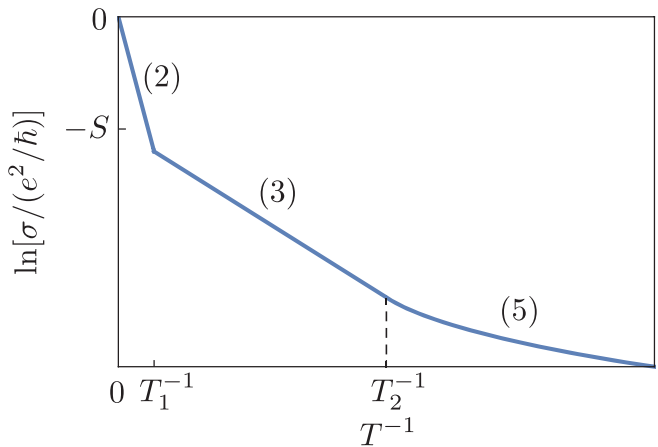}
    \caption{Schematic plot of the logarithm of the dimensionless conductivity $\sigma/(e^2/\hbar)$ as a function of the inverse temperature $T^{-1}$. At high temperature $T > T_1 = \Delta/S$, the conductivity has activation energy $\Delta$. At intermediate temperature $ T_2 < T < T_1$ where $T_2 = E_C^2/T_{\rm ES}$, the conductivity is dominated by activated hopping (AH). At low temperature such that $T < T_2$, AH is replaced by the ES law. Numbers adjacent to different parts of the line show corresponding equations. }
    \label{fig:conductivity}
\end{figure}

In situations where the typical tunneling transparency $P=\exp(-S)$
of the insulating barrier separating neighboring puddles is small (the action $S$ in units of $\hbar$ is large), one can envision a sequence of three mechanisms of activated transport replacing each other with decreasing temperature, as in a lightly doped wide gap semiconductor~\cite{shklovskii1984}.
This three-mechanism sequence is illustrated in Fig.~\ref{fig:conductivity}.
At relatively large temperature $T$ electrons and holes can be activated from the chemical potential to the percolation level (i.e., the classical mobility edge). As we show in detail in the Appendix \ref{app:percolation}, the activation energy for this process is exactly equal to $\Delta$ when the chemical potential $\mu = 0$ (i.e., at the charge neutrality point).
Thus, the conductivity at such large temperatures is given by 
\begin{align}  \label{eq:sigmaD}
    \sigma = \sigma_1 \exp(-\Delta/T), \hspace{5mm} (T_1 \ll T \ll \Delta) 
\end{align}
with the prefactor $\sigma_1 \sim e^2/\hbar$. Here and everywhere in this paper we use energy units for the temperature $T$ (absorbing $k_B$ in its definition).

At lower temperatures, this mechanism yields to hopping of electrons between electron and hole puddles near the chemical potential.
Similarly to the case of granular metals~\cite{chen2012, zhang2004}, the activation energy of such hopping is first determined by the typical puddle charging energy $E_C$,
\begin{align} \label{eq:sigma_AH}
\sigma=\sigma_2 \exp(-E_C/T), \hspace{5mm} (T_2 \ll T \ll T_1).
\end{align}
Here the prefactor $\sigma_2\sim (e^2/\hbar) \exp(-S) \ll (e^2/\hbar)$. We show below that 
\begin{align} \label{eq:ec_2}
E_C =\alpha^2 \Delta (\Delta/\Gamma)^{4/3} \ll \Delta. 
\end{align}
Here, $\alpha = e^2/(\kappa \hbar v)$ is the analog of the fine structure constant and $\kappa$ is the dielectric constant of the substrate. With the standard semiconductor value $v\sim 10^{6}$ m s$^{-1}$; and with $\kappa=4$ for SiO$_2$, 11 for insulating GaAs, 20 for HfO$_2$ and 1000 for PbTe; $\alpha$ can vary from 1 to $10^{-3}$. Below, in our theory, we use $\alpha$ as a small parameter, $\alpha \ll 1$.

At even lower temperatures, the activated hopping (AH) crosses over to the Efros-Shklovskii (ES) law,
\begin{align} \label{eq:sigma_ES}
\sigma=\sigma_3 \exp[-\left(T_{\rm ES}/T\right)^{1/2}], \hspace{5mm} (T \ll T_2),
\end{align}
with $\sigma_{3} \sim e^2/\hbar$. 
We show below that in this regime,
\begin{align} \label{eq:tes}
T_{\rm ES} = \alpha \Delta (\Delta/\Gamma)^{37/9} \ll \Delta,
\end{align}
and the temperatures associated with the crossover between the different regimes are 
\begin{align} \label{eq:t_1}
T_1 = \alpha\Delta(\Gamma/\Delta)^{34/9},
\end{align}
\begin{align} \label{eq:t_2}
T_2 = \alpha^3 \Delta (\Gamma/\Delta)^{13/9}.
\end{align}
Above we assumed that metallic gates are far enough from the semiconductor so that there is no screening of electron-hole Coulomb interaction leading to a crossover between the ES and Mott law of variable-range hopping. Such crossover happens when the hop length becomes larger than the distance to the gate~\cite{van_keuls1997,bennaceur2012}.

Similar results for all three mechanisms of conductivity are obtained below for thin films of 3D topological insulators. 
However, such a three-mechanism sequence is not observed in most experiments~\cite{Olshanetsky2015, Kvon2020, Pan2020, zou2010, taychatanapat2010, Serlin_Intrinsic_2020, Stepanov_untying_2020, Park_Flavour_2021, cao_correlated_2018, Cao_unconventional_2018, Xie_spectroscopic_2019, choi2021interactiondriven,  nandi2018,chong2021,checkelsky2012,chang2013,he2013,mogi2015,liguozhang2017,wang2018,fox2018,moon2019,rosen2019,rodenbach2021,fijalkowski2021}. 
Instead, experiments tend to report an activated conductivity with activation energy much smaller than $\Delta$.

Here we suggest a possible explanation for such low activation energies. We show that at $\Gamma/\Delta > (\Gamma/\Delta)_c$, electrons are not localized in single puddles and the first two regimes of conductivity are absent. For the general case presented above, $(\Gamma/\Delta)_c =\alpha^{-9/41}$. 
The only remaining mechanism is the ES law with very small $T_{\rm ES}$. This means that the low-temperature ``local activation energy'' is much smaller than $\Delta$. 
Such a theory predicts that the prefactor should be close to $e^2/\hbar$. 
This prediction agrees with some experiments~\cite{fox2018,rodenbach2021}, but disagrees with others~\cite{rosen2019,fijalkowski2021}.

\section{Fractal geometry of puddles} 
\label{sec:fractal}

In this section, we briefly review some geometrical fractal properties of 2D puddles at $\Gamma \gg \Delta$~\cite{isichenko1992}.
The characteristic size (diameter) of a puddle is given by 
\begin{equation}\label{eq:a}
    a = \lambda (\Gamma / \Delta)^{\nu},
\end{equation}
where $\nu = 4/3$ and $\lambda$ is the electron screening radius. The perimeter of a puddle is
\begin{align} \label{eq:L}
    L = a (\Gamma / \Delta) = \lambda (\Gamma / \Delta)^{\nu + 1}.
\end{align}
The perimeter $L$ is parametrically longer than the diameter $a$ because puddles have many ``fingers,'' which are interlocked with other fingers of neighboring puddles (see Fig.~\ref{fig:fingers}).
The area of a puddle is given by
\begin{align}\label{eq:A}
    A = \lambda^2 (\Gamma/\Delta)^{2\nu - \beta},
\end{align}
where $\beta = 5/36$.
The typical separation distance between nearest-neighbor puddles is
\begin{align}\label{eq:w}
    w = \lambda \Delta /\Gamma.
\end{align}

In order to estimate $\Gamma$ and $\lambda$, we can use the self-consistent theory of Ref.~\onlinecite{skinner2013a}, which dealt with the disorder potential at the surface of a bulk TI created by charged impurities with 3D concentration $N$. 
In our case, the substrate plays the role of the TI bulk and the two-dimensional semiconductor plays the role of the TI surface.
The band gap $\Delta$ that exists in our case is not important for determining the values of $\Gamma$ and $\lambda$ when $\Gamma \gg \Delta$. 
To begin, we relate $\Gamma$ to $\lambda$ as the typical Coulomb energy created by charge fluctuations in a volume $\lambda^3$:
\begin{align}\label{eq:gamma0}
\Gamma = \frac{e^2}{\kappa\lambda} (N\lambda^3)^{1/2}.
\end{align}
This relation leads to a typical 2D density of states,
\begin{align}\label{eq:dos}
    g = \kappa^2 \alpha^2 \Gamma/e^4,
\end{align}
which in turn leads to the screening radius,
\begin{align}\label{eq:lambda1}
\lambda = \frac{\kappa}{e^2 g} = \frac{e^2}{\alpha^2\kappa \Gamma}.
\end{align}
Solving Eqs.~\eqref{eq:gamma0} and \eqref{eq:lambda1} for $\Gamma$ and $\lambda$, we get~\cite{skinner2013a}
\begin{align}\label{eq:gamma}
    \Gamma & = \frac{e^2 N^{1/3}}{\kappa \alpha^{2/3}}, \\
    \lambda & = \alpha^{-4/3} N^{-1/3}.\label{eq:lambda}
\end{align}

\section{Tunneling action, localization length and conductance}
\label{sec:tunneling} 

Let us now calculate the hopping conductivity of the system of fractal metallic puddles separated by narrow insulating gaps shown in Fig.~\ref{fig:puddles}. 
In this section, we estimate the dimensionless tunneling action $S$.  
The value of $S$ is determined by the tunneling length $r=\Delta/eE$ in the spatially varying electric field $E$ created by impurities,
\begin{align}\label{eq:action}
S = \frac{r \Delta}{\hbar v} = \frac{\Delta^2}{eE\hbar v}.
\end{align}
It is tempting to use $\Gamma/e\lambda$ for $E$ and arrive at $S = \alpha^{-1} (\Delta/\Gamma)^2$.
However, the electric field has strong fluctuations at short distances, so the typical electric field depends on the tunneling distance $r$. Since a cube of size $r$ has a typical excess impurity charge $\sqrt{Nr^3}$, the typical electric field associated with the length scale $r$ is $E(r)= e(Nr^3)^{1/2}/\kappa r^2$, which grows with decreasing $r$. 
Also, due to the large perimeter length $L$ of puddles, we can find rare places where the random electric field is created by a larger-than-average number of excessive charges, $M\gg (Nr^3)^{1/2}$, leading to even larger electric field $E(r)= e M/\kappa r^2$. Below we find the optimal values of $M$ and $r$ which determine $S$, and we arrive at a value of $S$ much smaller than 
the naive estimate quoted above. 
Our optimization procedure is a mesoscopic version of the optimization used in the theory of the interband absorption of light in compensated three-dimensional semiconductors~\cite{shklovskii1984,shklovskii1970}. It is also similar to the theory of fluctuation-induced excess currents in reverse biased $p$-$n$ junctions~\cite{raikh1985}. 

Below we use $S$ to calculate the localization length $\xi$ that determines hopping transport. Thus, we are interested in fluctuations of the electric field which, although rare, happen roughly once at every interface between nearest-neighboring puddles. Thus,
\begin{align}\label{eq:m}
    (L/\lambda) \exp[-\frac{M^2}{Nr^3}] = 1.
\end{align}
Here we use the Gaussian probability of finding net charge $M$
in a cube of size $r$. For tunneling across the gap $2\Delta$, we need the potential difference across the cube $Me^2/\kappa r=\Delta$. In other words, $r=r(M) = M e^2 /\kappa\Delta$. Substituting $r(M)$ into Eq.~\eqref{eq:m} and solving for $M$ gives
\begin{align}
    M = \frac{\alpha^{-2} (\Delta/\Gamma)^3}{\ln[(\Gamma/\Delta)^{7/3}]}, 
\end{align}
which at $\Gamma \gg \Delta$ corresponds to $r(M) \ll w \ll \lambda$.

Substituting the electric field $E = M e / \kappa r^{2}(M)$ into the tunneling action given by Eq.~\eqref{eq:action}, we have
\begin{align}\label{eq:action_meso}
    S = \frac{\alpha^{-1} (\Delta/\Gamma)^3}{\ln[(\Gamma/\Delta)^{7/3}]} \simeq \alpha^{-1} (\Delta/\Gamma)^{34/9}.
\end{align}
In the last step, we used the power-law approximation $\ln x \approx x^{1/3}$, which is valid for $x \in (3,100)$ with accuracy better than 30\% . 

Now we can calculate the electron localization length $\xi$, which we need below to calculate the hopping conductivity. 
Consider a tunneling path with a displacement $x \gg a$. The action associated with this path is dominated by the action for tunneling across the narrow insulating gaps between puddles, which the electron must cross every time it displaces across one puddle diameter $\xi$. 
Consequently, the total action of the tunneling path is $Sx/a = x/\xi$, where 
\begin{align}\label{eq:xi} 
    \xi = a/S = \alpha a (\Gamma/\Delta)^{34/9}.
\end{align}

Notice, however, that the fast decrease of $S$ with growing $\Gamma/\Delta$ leads to a fast increase of the dimensionless conductance between two neighboring puddles,
\begin{align}\label{eq:g} 
 G = (L/\lambda)\exp(-S),
\end{align}
so that we get $G=1$ at some critical value $(\Gamma/\Delta)_c$. 
Substituting Eq.~\eqref{eq:action_meso} into Eq.~\eqref{eq:g} and setting $G=1$, we arrive at an estimate for the critical disorder strength~\footnote{In the limit of $\alpha \to 0$, the asymptotic expression to first order reads $(\Gamma/\Delta)_c = \alpha^{-1/3}[\ln(\alpha^{-1})]^{-2/3}$.},
\begin{align} \label{eq:critical} 
    (\Gamma/\Delta)_c = \alpha^{-9/41},
\end{align}
valid for $\alpha \in (1.2 \times 10^{-4}, 0.12)$.
This range of $\alpha$ is obtained by substituting Eq.~\eqref{eq:critical} into the requirement for the argument of the logarithm $(\Gamma/\Delta)^{7/3} \in (3, 100)$~\footnote{Note that this mesoscopic optimization method based on Eq.~\eqref{eq:m} is self-consistent if $\Gamma/\Delta < (\Gamma/\Delta)_c$ (or $G < 1$), so that $e^S > L/\lambda$.}.

At larger $\Gamma / \Delta$, the localization length grows exponentially as $\xi = a e^G$.
This growth leads to a dramatic growth of the conductivity, which we dub an ``insulator--almost metal transition'' (IAMT), if the sample size is much larger than $\xi$. 
For a sufficiently small sample,  $(\Gamma/\Delta)_c$ effectively plays the role of disorder-induced insulator-metal transition.

\section{Hopping conductivity}
\label{sec:hopping}

At moderate disorder when $1 < \Gamma/\Delta < (\Gamma/\Delta)_c$, electrons are well localized within a single puddle and the temperature dependence of the conductivity follows the three-mechanism sequence described above,
for which with decreasing temperature the activated conductivity with activation energy $\Delta$ is replaced first by AH and then by the ES law. 
In this case, the system is similar to a network of densely packed  metallic granules separated by a thin insulating matrix with Coulomb impurities, and we can follow the calculation of their conductivity~\cite{zhang2004,chen2012}.  

\begin{figure}[t]
    \centering
    \includegraphics[width=0.8 \linewidth]{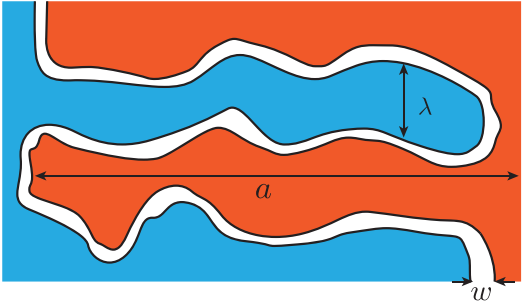}
    \caption{Schematic picture of interlocked ``fingers'' of neighboring puddles. Here the length of ``fingers'' $a$ is of the order of the puddle diameter. One can imagine that Fig.~\ref{fig:puddles} shows a vertical cross section of Fig.~\ref{fig:fingers}.} 
    \label{fig:fingers}
\end{figure}

Let us start from the discussion of AH conductivity at $\Delta/T \gg S$ or $T \ll T_1 = \Delta /S$. [One arrives at  Eq.~\eqref{eq:t_1} for $T_1$ with the help of Eq.~\eqref{eq:action_meso}.] In this case, the charging energy of a puddle $E_C$ replaces $\Delta$ as the activation energy for conductivity. 
In the case of large $\Gamma /\Delta$, we study the fractal structure of puddles which leads to a peculiar expression for $E_C$, smaller than the standard expression 
$E_C = e^2/\kappa a$. 
Namely, we are going to show that
\begin{align}\label{eq:newec}
E_C = \frac{e^2 \Delta}{\kappa a \Gamma }.
\end{align}
By substituting Eqs.~\eqref{eq:a} and \eqref{eq:lambda1} into Eq.~\eqref{eq:newec}, one arrives at Eq.~\eqref{eq:ec_2}.
Let us illustrate how this happens by comparing the self-capacitance of an isolated puddle, $C_{0} \sim \kappa a$, with the capacitance $C$ of the same puddle surrounded by other puddles.
In the latter case, because our puddle has metallic properties, an excess electron charge $e$ spreads to the border (perimeter), 
while neighboring metallic puddles provide opposite charge on the other side of the border.
Thus, all of the electric field is concentrated at the border between two puddles, mostly between the long fingers of the electron and hole puddles shown in Fig.~\ref{fig:fingers}.

This mechanism of enhanced capacitance was recognized by the electrical engineering community~\cite{samavati1998}.
In our system it means that $C \sim \kappa L$, and $E_C = e^2/C$ leads to Eq.~\eqref{eq:newec}.

The use of the activation energy $E_C$ is justified when it is larger than the energy level spacing in a puddle.
The level spacing is given by
\begin{align}
    \delta = (gA)^{-1} = \alpha^2 (\Delta / \Gamma)^{55/36} \Delta,
\end{align}
where $g$ is the 2D density of states (DOS) given by Eq.~\eqref{eq:dos}, and $A$ is the area of a puddle given by Eq.~\eqref{eq:A}.
Therefore, the ratio $\delta / E_C = (\Delta/\Gamma)^{7/36} \ll 1$ and our use of $E_C$ is legitimate.

Let us now consider the ES conductivity which replaces AH at low temperatures.
In the ground state, each puddle $i$ of our system is charged by a random fractional charge $|q_{i}| \leq e/2$. 
This charging happens because some impurities contribute their potential to neighboring puddles effectively by sharing their charge between neighboring puddles, so that each puddle effectively gets a fraction of impurity charge $e$. 
On the other hand, electrons contribute their integer charge $e$ to their puddles. Fractional charging provides background disorder and creates a random potential that results in localized electron states and enables the formation of the Coulomb gap around the chemical potential~\cite{zhang2004,chen2012}.
This Coulomb gap leads again to conductivity described by the ES law in the low-temperature limit.

We can calculate the constant $T_{\text{ES}}$ in the ES law starting from the standard expression $T_{\text{ES}}  = e^2/\kappa \xi$~\cite{efros1975,shklovskii1984}. 
Using  Eq.~\eqref{eq:xi} for $\xi$, we arrive at Eq.~\eqref{eq:tes}.
We see now that $T_{\text{ES}} \ll \Delta$. 
Equating $(T_{\text{ES}}/T)^{1/2}$ to $E_C/T$ with the help of Eq.~\eqref{eq:tes}, 
we arrive at the expression of $T_2$ as Eq.~\eqref{eq:t_2}.

\section{Thin film of 3D topological insulator}
\label{sec:TI}

In previous sections, we dealt with a general model of a trivial 2D semiconductor with gapped Dirac spectrum, given by Eq.~\eqref{eq:spectrum}. 
In this section, we concentrate on the special case of a thin film of 3D TI, where the narrow gap $2\Delta$ can be a result of the hybridization of surface states on opposite surfaces of the film~\cite{liu2010,lu2010,linder2009,zhang2010a,sakamoto2010,zhang2013,kim2013,nandi2018,chong2021} or may be created by a finite concentration of magnetic dopants such as Cr~\cite{chen2010,xu2012,checkelsky2012,chang2013,he2013,mogi2015,liguozhang2017,wang2018,fox2018,moon2019,rosen2019,rodenbach2021,fijalkowski2021}. 
Because of the promise of such films to achieve metrological precision of the quantum anomalous Hall effect~\cite{yu2010,chang2013,jinsongzhang2013,mogi2015,fox2018,rodenbach2021,fijalkowski2021} and the quantum spin Hall effect~\cite{liu2010,lu2010,linder2009,chong2021},
in this section we are more specific with material parameters and numerical estimates.

We have in mind TI thin films based on (Bi$_{x}$ Sb$_{1-x}$)$_2$Te$_{3}$~\cite{nandi2018}, or (Bi$_{x}$ Sb$_{1-x}$)$_2$ (Te$_{y}$ Se$_{1-y}$)$_3$~\cite{chong2021} which have very large dielectric constant $\kappa \sim 200$~\cite{richter1977,borgwardt2016,bomerich2017}. 
Using $\kappa \sim 200$ and the Fermi velocity of TI $v \sim 4 \times 10^5$ m/s~\cite{jszhang2011}, one gets $\alpha \sim 0.027$.
We assume that such a film of width $d \sim 7$ nm is deposited on a substrate with much smaller dielectric constant $\kappa_e \ll \kappa$, so that the electric field created by Coulomb impurities residing inside the film is trapped within the film~\cite{rytova1967,chaplik1971,keldysh1979,huang2021a}. 
This trapping slows down the decay of the Coulomb potential with distance and enhances the role of impurities. 
TI films typically have a large ($N \sim 10^{19}$ cm$^{-3}$) concentration of Coulomb impurities, which allows us to study only their effect and to assume that impurities inside the substrate play no role. 

As in the previous section, we calculate the tunneling action $S$ and the critical ratio $(\Gamma/\Delta)_c$, and we describe the hopping conductivity of the film. 
According to Ref.~\cite{huang2021a}, due to the peculiar electrostatics of TI films, the expression of $\Gamma$ is the same as Eq.~\eqref{eq:gamma}, while the expression for $\lambda$ becomes
\begin{align}\label{eq:lambda_ti}
    \lambda = \alpha^{-2/3} (Nd^3)^{-1/6} d,
\end{align}
which is valid if $\lambda > d$. 
Using the above estimates of $\alpha$, $d$, and $N$, we get $\lambda \sim 50$ nm, so that this estimate is valid.

Notice that the electric field in the plane of the film created by charge fluctuations in a disk of radius $r$ and thickness $d$ is given by
\begin{align}\label{eq:E_ti}
    E = \frac{e \sqrt{Nr^2 d}}{\kappa r d} = \frac{e}{\kappa} \sqrt{\frac{N}{d}},
\end{align}
which turns out to be independent of $r$.
Therefore, there is no enhancement of the electric field at scales shorter than $\lambda$, as there was in Sec.~\ref{sec:tunneling}. 
Substituting Eq.~\eqref{eq:E_ti} (or, equivalently, $E=\Gamma/e\lambda$) into Eq.~\eqref{eq:action}, we arrive at the action
\begin{align}\label{eq:action_w_ti}
    S = \frac{w \Delta}{\hbar v} = \frac{\lambda  \Delta^2}{\hbar v \Gamma} =  \alpha^{-1/3} (Nd^3)^{1/6} (\Delta / \Gamma)^{2}
\end{align}
for tunneling between neighboring puddles.
However, the electric field $E = eM/\kappa \lambda d$ can still be enhanced by a rare fluctuation of the number of charges $M \gg (N\lambda^2 d)^{1/2}$ with
Gaussian probability $\exp(-M^2/N\lambda^2 d)$. This replaces Eq.~\eqref{eq:m} by
\begin{align}
    (L/\lambda) \exp[-\frac{M^2}{N\lambda^2d}] = 1.
\end{align}
Solving the above equation, we obtain the largest $M$ available in the perimeter
\begin{align}
    M = \qty{N\lambda^2 d \ln[(\Gamma/\Delta)^{7/3}]}^{1/2}.
\end{align}
Now substituting the electric field $E = eM/\kappa \lambda d$ into the action Eq.~\eqref{eq:action} gives
\begin{align}\label{eq:action_meso_ti}
    S = \frac{\alpha^{-1/3} (N d^3)^{1/6} (\Delta / \Gamma)^{2}}{\qty{\ln[(\Gamma/\Delta)^{7/3}]}^{1/2}} \simeq \alpha^{-1/3} (N d^3)^{1/6} (\Delta / \Gamma)^{43/18},
\end{align}
which is smaller than the action given by Eq.~\eqref{eq:action_w_ti}. In the last step, as in Sec.~\ref{sec:tunneling}, we used the power-law approximation $\ln x=x^{1/3}$ valid for $x \in (3,100)$ with accuracy better than 30\%.

Substituting Eq.~\eqref{eq:action_meso_ti} into the expression of $G$, given by Eq.~\eqref{eq:g}, and setting $G=1$, we arrive at the critical point~\footnote{In the limit of $\alpha (N d^3)^{-1/2}\to 0$, the asymptotic expression to first order reads $(\Gamma/\Delta)_c = \alpha^{-1/6} (N d^3)^{1/12}\qty{\ln[\alpha^{-1} (N d^3)^{1/2}]}^{-3/4}$.},
\begin{align} \label{eq:critical_ti} 
    (\Gamma/\Delta)_c = \alpha^{-2/19}(N d^3)^{1/19}.
\end{align}
Using the estimates $\alpha \sim 0.027$, $N \sim 10^{19}$ cm$^{-3}$ and $d=7$ nm, we get $(\Gamma/\Delta)_c = 1.6$.

Let us switch to the hopping conductivity of the thin TI film and start from 
the charging energy of a puddle.
Similar to Sec.~\ref{sec:fractal}, the capacitance of a puddle within the film is determined by the long border of the puddle with neighboring puddles.
Near the border, there are two stripes of length $L$ and width $\lambda$ with charges $-e$ and $e$. But now, the electric field at the border is concentrated in the film of width $d\ll\lambda$ with the large dielectric constant $\kappa$. This changes the capacitance of the puddle border to $C\sim \kappa L(d/\lambda)$ and leads to 
\begin{align}\label{eq:ec_ti}
    E_C =  (e^2/\kappa d)(\lambda/L) = (e^2/\kappa d)(\Delta / \Gamma)^{7/3},
\end{align}

At lower temperatures $T<T_2$, the conductivity obeys the ES law with the characteristic temperature $T_{\rm ES} = e^2/\kappa_e \xi$, where $\xi = a/S$. 
Note that here we use $\kappa_e$ because, at large distances, the electric field lines leave the film and go through the environment. Using Eqs.~\eqref{eq:a} and \eqref{eq:action_meso_ti}, we get
\begin{align} \label{eq:tes_ti}
    T_{\rm ES} = \alpha(\kappa/\kappa_e)\Delta (\Delta/\Gamma)^{49/18}.
\end{align}

Equations.~\eqref{eq:tes_ti} and ~\eqref{eq:critical_ti} show that in TI films, as in trivial semiconductors [cf. Eqs.~\eqref{eq:tes} and ~\eqref{eq:critical}],
a reduction of $T_{\rm ES}$ and a crossover from strong localization to practically metallic conductivity happens dramatically quickly when $\Gamma$ exceeds $\Delta$.

We now estimate the characteristic energies $\Gamma$, $\Delta$, $E_C$, $T_{\rm ES}$, $T_1$, and $T_2$ for TI thin films based on (Bi$_{x}$ Sb$_{1-x}$)$_2$ (Te$_{y}$ Se$_{1-y}$)$_3$. Using $\kappa=200$, $\alpha = 0.027$, and $N = 10^{19}$ cm$^{-3}$, we have $\Gamma \simeq 17$ meV. 
The hybridization gap is related to the thickness by $\Delta = \Delta_0 e^{-d/d_0}$ with $\Delta_0 = 0.5$ eV and $d_0=2$ nm~\cite{chong2021}. 
For example, if $d = 7$ nm, then $\Gamma \simeq 17$ meV, $\Delta \simeq 15$ meV, $S \simeq 3$, $E_C \simeq 0.8$ meV, and $T_{\rm ES} \simeq 130$ K (here assume that the film has boron nitride on both sides and use $\kappa_e =5$). Temperatures $T_1 = \Delta/S \simeq 60$ K and $T_2 = E_C^2/T_{\rm ES} = 0.6$ K are obtained by equating Eqs.~\eqref{eq:sigmaD} to \eqref{eq:sigma_AH}, and \eqref{eq:sigma_AH} to \eqref{eq:sigma_ES}, respectively. In this case, apparently ES conductivity starts 
when $(T_{\rm ES}/T_2)^{1/2} \sim 15$ so that the ES law is hardly observable because of very large resistance. Thus, observable activation energy is given by $E_C \sim 0.05 \Delta$. 

In slightly thicker films with $d\geq 8$ nm, the half gap $\Delta(d) \leq 9$ meV and $\Gamma/\Delta > (\Gamma/\Delta)_c$, so that they are almost metallic and show ES conductivity with much smaller $T_{\rm ES}$. On the other hand, one can show that in thinner films, $d < 7$ nm, for which $\Delta > \Gamma $ activation energy $E_c= e^2/\kappa d$ and $T_{\rm ES}=\alpha (\kappa/\kappa_e) \Delta$ so that practically, conductivity is similar to films with $d = 7$ nm.
Notice that critical thickness $d=d_c =7$ nm is very sensitive to values of 
$\Delta_0$, $N$, $\kappa$, $\alpha$, and, most importantly, $d_0$, which are different for different materials. 
This can explain the differences between experimental results in Refs.~\onlinecite{nandi2018,chong2021}.

For the case of magnetically doped TI thin films, the exchange half gap $\Delta$ induced by magnetic impurities is not directly related to $d$ and is of the order of 20 meV~\cite{rodenbach2021,lu2021,fijalkowski2021}, so that we have practically the same numbers as in the previous example.

\section{Summary and Conclusion}
\label{sec:conclusion}

In this paper, we have considered the temperature-dependent conductivity of a two-dimensional insulator subjected to the random potential created by Coulomb impurities in the substrate. Our primary results can be summarized as follows. First, the random potential of charged impurities necessarily produces large band bending. We focus here on the case where the impurity concentration is large enough that the disorder potential $\Gamma \gg \Delta$, and the system can be described as a network of large and closely spaced fractal puddles  (Fig.~\ref{fig:puddles}) separated by narrow insulating barriers (Fig.~\ref{fig:fingers}). This case is characterized by the ``three-mechanism sequence'' of temperature-dependent conductivity illustrated in Fig.~\ref{fig:conductivity}. Only the highest-temperature regime has an activation energy $E_a$ equal to half the band gap $\Delta$. The middle regime, with activated hopping between puddles (AH), exhibits a parametrically smaller activation energy whose value depends on the impurity concentration, while the lowest-temperature regime corresponds to Efros-Shklovskii conductivity, which may appear as an even smaller activation energy when measured over a limited temperature range.

Second, when the impurity concentration $N$ exceeds some critical value, the tunnel barriers between puddles become thin enough to be nearly transparent, and electrons are delocalized across many puddles. In this limit, the localization length grows exponentially with increased disorder, and the corresponding activation energy falls exponentially, so that in mesoscopic samples one effectively has a disorder-induced insulator-to-metal transition. 

Our results have implications for a wide variety of experiments on 2D electron systems with a narrow energy gap. Some of these include 2D and thin 3D TIs, Bernal bilayer graphene with a perpendicular displacement field, and twisted bilayer graphene, as mentioned in Sec.~\ref{sec:intro}. In such systems, the temperature-dependent conductivity is often used as a primary way to diagnose the magnitude of energy gaps. Our results here suggest that such studies suffer an essentially unavoidable limitation, since the apparent activation energy $E_a$ at low temperature has no simple relation to the energy gap, and, in general, $E_a$ can be taken only as a weak lower bound. No wonder that the transport activation energy in many cases is 100 times smaller than the value expected theoretically or measured by probes such as optical absorption or tunneling spectroscopy.

In principle, one can infer the band gap by measuring the activation energy at the highest-temperature regime. However, the existence of this regime practically requires a low enough disorder that electron and hole puddles are small and well separated from each other. Even in this case, experiments using transport to estimate $\Delta$ should first demonstrate two distinct regimes of constant activation energy, and then use only the value from the higher-$T$ regime as an estimate of $\Delta$.

The existence of an apparent disorder-induced IMT at $\Gamma/\Delta > (\Gamma/\Delta)_c$ is an especially striking result of our analysis. For conventional insulators, this apparent transition cannot be called a true IMT, since in 2D the zero-temperature conductance flows toward zero in the thermodynamic limit for any finite amount of disorder \cite{Abrahams1979}. However, the situation may be different for thin TI films, since the spin-orbit coupling of the TI surface states permits a stable metallic phase \cite{Hikami1980, Mong2012}. A full theory of this IMT in TI films is beyond the scope of our current analysis.


\begin{acknowledgments}
We are grateful to Stevan Nadj-Perge, Koji Muraki, Shahal Ilani, Ilya Gruzberg, and David Goldhaber-Gordon for helpful conversations.
Y.H. is supported by the William I. Fine Theoretical Physics Institute. B.S. was partly supported by NSF Grant No. DMR-2045742.
\end{acknowledgments}

\appendix

\section{Activation to the classical mobility edge}
\label{app:percolation}

\subsection{Percolation-based argument for $E_a = \Delta$}

In this appendix, we consider the process of thermal activation of electrons from the chemical potential to the classical mobility edge. We show that, generically, this process leads to an activation energy
\be 
E_a = |E_\textrm{be} - \mu|,
\label{eq:Eamob}
\ee 
where $E_\textrm{be} = \pm \Delta$ represents the mean energy of the nearest band edge (conduction band or valence band) and $\mu$ is the chemical potential, both defined relative to the mid gap~\cite{mahmoodian2020}.  In the maximally insulating state, where $\mu = 0$, this equation gives
\be 
E_a = \Delta,
\label{eq:Eadelta}
\ee 
as in the non disordered system.

To show this result, we first consider the case $\mu = 0$. In this case, the mean energy of the conduction-band edge is $+\Delta$ and the mean energy of the valence-band edge is $-\Delta$, although the disorder potential causes both band edges to wander in energy as a function of position (see Fig.~\ref{fig:puddles}.)  Let us now consider the process of drawing spatial contours of constant energy $E$ for one of the two band edges (say, the conduction band).  The contours corresponding to $E = 0$ are small closed curves which surround electron puddles.  As the energy $E$ is increased, these contours grow in diameter and an increasing fraction of the system's area is inscribed within such contours. This inscribed area corresponds to the regions of the system that are accessible to conduction-band electrons with energy $E$.

At some point, as the contour energy $E$ is increased, the area inscribed within contours becomes large enough that it comprises half of the total area of the system. At this energy $E_p$, there is a percolation transition, such that a conduction band electron with $E > E_p$ can move freely across a macroscopic distance and contribute to bulk conduction. The energy $E_p$ is therefore equal to the position of the classical mobility edge for the conduction band, and the activation energy for this process $E_a = E_p$. That percolation occurs when half the area of the system is encompassed by contours can be argued simply on the basis of symmetry: in two dimensions, it is not possible for a continuous percolation cluster and its complement to percolate simultaneously. Thus the critical area fraction for a continuous and symmetric random potential is equal to $1/2$ \cite{shklovskii1984}.

Thus, the energy of activation to the conduction band is equal to the energy at which half of the system's area is accessible via conduction-band states. For a symmetric potential, this energy corresponds exactly to the mean energy of the conduction band, and therefore $E_a = \Delta$, as announced above. 

This same argument can easily be extended to the case where the chemical potential is not at zero, but is instead at some finite energy $\mu$ relative to the mean position of the mid gap. The percolation level $E_p$ remains unchanged relative to the mid gap, and thus the energy of activation to the conduction band is $\Delta - \mu$. The energy of activation to the valence band is $\Delta + \mu$.

Thus, the activation energy for activation to the nearest mobility edge is identical to what it would have been in the non disordered case. Its origin, however, is nontrivial, and involves a symmetry of percolation in a two-dimensional continuous potential. In three dimensions, for example, there is no such symmetry since a given space and its complement may percolate simultaneously. Consequently, the activation energy for percolation is significantly smaller than the difference in energy between the chemical potential and the mean position of the closest band edge, $E_a \approx 0.3 |E_\textrm{be} - \mu|$ \cite{skinner2012, Chen2016}.

\subsection{Computer modeling}

In order to confirm Eq.~(\ref{eq:Eamob}), we consider a simple computer model of a 2D electron system of dimensions surrounded by a three-dimensional environment containing charged impurities.  We simulate the electron system as a square grid of $L_0 \times L_0$ discrete points, embedded as the midplane of a cube containing $L_0^3$ charged impurities. We define our coordinates such that the 2D electron system comprises the $x-y$ plane, while impurities are uniformly and randomly distributed within the region $-L_0/2 < x,y,z < L_0/2$ (impurities are not constrained to reside at integer values of $x,y,z$). A given discrete point $i$ in the $x$-$y$ plane may have charge $q_i = -1, 0, 1$, with the values corresponding to the location of the chemical potential relative to the conduction band and valence-band edges at site $i$. Specifically, $q_i = -1$ corresponds to the chemical potential being above the conduction-band edge, $q_i = +1$ indicates that the chemical potential is below the valence-band edge, and $q_i = 0$ if the chemical potential is within the band gap.
\begin{figure}[t]
\begin{center}
\includegraphics[width=0.8\columnwidth]{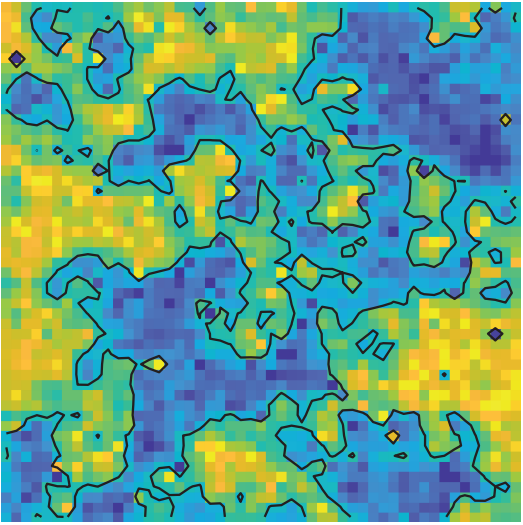}
\end{center}
\caption{An example of the energy $E_\textrm{CB}$ of the conduction band edge as a function of position, as given by our numerical simulation. Light (yellow) colors indicate high energy, and dark (blue) colors correspond to low energy. The black contour indicates the energy of the percolation level, $E_\textrm{CB} = \Delta$.  This example corresponds to $\Delta = 5, L_0 = 50$.}
\label{fig:potentialexample}
\end{figure}
The corresponding semiclassical Hamiltonian for this system is
\be 
H = \sum_i \phi_i^\textrm{imp} q_i - \Delta \sum_i q_i + \sum_{i, j \neq i} \frac{q_i q_j}{r_{ij}}
\ee 
where $r_{ij}$ is the distance between the sites $i$ and $j$ in the $x$-$y$ plane. The quantity $\phi_i^\textrm{imp}$ represents the potential created by bulk impurities at site $i$, and is given by
\be 
\phi_i^\textrm{imp} = \sum_{k} \frac{q_k^\textrm{imp}}{r_{ik}}.
\ee 
Here, $r_{ik}$ represents the distance between the bulk impurity $k$ and the site $i$; the impurity charge is $q_k^\textrm{imp} = \pm 1$. The corresponding energies of the conduction-band and valence-band edges at site $i$ are
\begin{align}
E_i^\textrm{CB} & = - \phi_i + \Delta \\
E_i^\textrm{VB} & = - \phi_i - \Delta,
\end{align}
where
\be 
\phi_i =  \phi_i^\textrm{imp}  + \sum_{j \neq i} \frac{q_j}{r_{ij}} 
\ee 
is the total electric potential at site $i$.
\begin{figure}[t]
\begin{center}
\includegraphics[width=0.95\columnwidth]{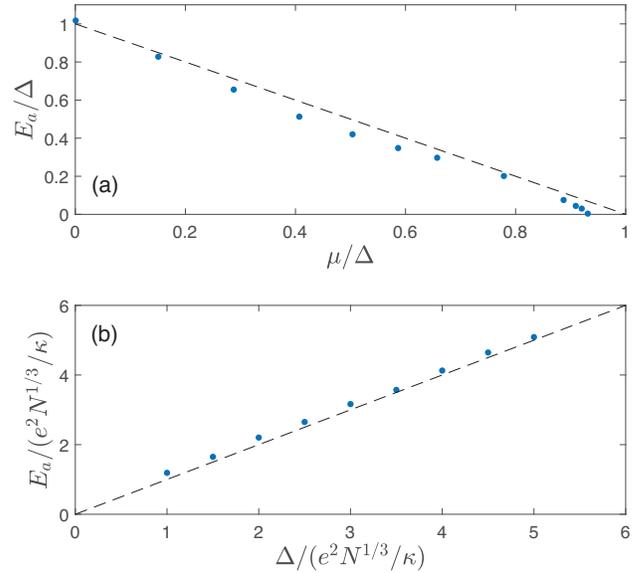}
\end{center}
\caption{Numerical results for the activation energy as a function of (a) the chemical potential $\mu$ and (b) the band gap $\Delta$. The dashed lines in (a) and (b) correspond to Eqs.~(\ref{eq:Eamob}) and (\ref{eq:Eadelta}), respectively. Both plots use a system size $L_0 = 50$.}
\label{fig:mobilityedge}
\end{figure}
The intent of this model is to describe the spatial meandering of the band edge for a 2D system surrounded by bulk impurities. Once the corresponding energies $E_i^\textrm{CB, VB}$ are known, we can calculate the energy $E_a$ that corresponds to the classical mobility edge for the nearest band. Specifically, $E_a$ corresponds to the minimal energy such that there exists a percolating path across opposite faces of the system using only lattice sites with $E_i^{CB} < E_a$. The model discretizes the electron system on a length scale given by the mean distance between bulk impurities. The corresponding correlation length of the random potential is given by
\be 
\Lambda = \frac{\Delta^2 \kappa^2}{2 \pi e^4  N},
\ee 
and so such discretization is unimportant when $\Delta / (e^2 N^{1/3} / \kappa )$ is large. We use nearest-neighbor percolation on the square lattice, but since the potential is correlated over length scales much larger than the lattice constant, this choice is unimportant in the limit of large $\Delta$.

In order to find the energies $E_i^\textrm{CB, VB}$, we first need a solution for the charges $q_i$ of each site $i$ in the ground state. Finding such a solution is a difficult numerical problem. We use the numerical algorithm described in Refs.~\cite{shklovskii1984, skinner2012, Chen2016} to find a pseudo-ground state that is minimized with respect to changing any one or two values of $q_i$ simultaneously. This algorithm is known to give a good approximation for the properties of the ground state.
The resulting solution for the charge $q_i$ of each site allows us to define the energies $E_i^\textrm{CB, VB}$, and thus to find the percolation level for both the conduction band (CB) and valence band (VB). We iterate this procedure over many random choices of the impurity positions, and average over all such iterations.

We also examine the dependence of the activation energy on the chemical potential $\mu$.  The chemical potential can be tuned by adjusting the net concentration of charged impurities, which mimics the effect of a gate voltage.  $\mu = 0$ corresponds (on average) to equal numbers $L_0^{3/2}$ of $+$ and $-$ impurities, since we are defining $\mu$ relative to the average energy of the mid-gap, which at site $i$ has an energy $- \phi_i$.  We can thus define $\mu$ for the system as 
\be 
\mu = - \frac{1}{L_0^2} \sum_i \phi_i.
\ee 
This chemical potential is calculated after the pseudo-ground state is determined.
We then report $E_a$ relative to this chemical potential.

Figure \ref{fig:potentialexample} shows an example plot of the energy of the CB bottom, $E^\textrm{CB}$, as a function of position.
Figure \ref{fig:mobilityedge} gives the calculated activation energy as a function of chemical potential and band gap. It closely matches Eq.\ \eqref{eq:Eamob}.


%

\end{document}